Testing the science/technology relationship by analysis of patent citations of scientific papers after decomposition of both science and technology

Fang Han[1, 2]*[†], Christopher L. Magee[1, 3][†]

[1]SUTD-MIT International Design Center, Massachusetts Institute of Technology, Cambridge, MA 02139, USA.
[2]School of Economics and Management, Beijing University of Posts and Telecommunications, Beijing, 100876, CHINA.
[3]Institute for Data, Systems and Society, Massachusetts Institute of Technology, Cambridge, MA 02139, USA.
*Correspondence to: fhanmail@163.com (F.H.)
[†]Both authors contributed equally to this work.

**Abstract**

The relationship of scientific knowledge development to technological development is widely recognized as one of the most important and complex aspects of technological evolution. This paper adds to our understanding of the relationship through use of a more rigorous structure for differentiating among technologies based upon technological domains (defined as consisting of the artifacts over time that fulfill a specific generic function using a specific body of technical knowledge). The key findings of the work are:

Firstly, a Pearson correlation of 0.564 is found between technological relatedness among technological domains based upon patents citing other patents and technological relatedness among technological domains based upon patents citing similar scientific papers (assessed through 176 scientific categories developed by ISI). This result indicates that a large portion (but not all) of technological relatedness is due to relatedness of the underlying scientific categories.

Secondly, the overall structure of the links found between scientific categories and technological domains is many-to-many rather than focused indicating a science-fostered mechanism for fairly broad "spillover":

Specific technological domains cite a wide variety of scientific categories (in 2010 the *average* number of scientific categories cited by the 44 individual technological domains studied was 36);

Some scientific categories are cited in a variety of domains (124 scientific categories are cited in 30% or more of the 44 domains over the period studied, 1976-2013).

Thirdly, Some evidence is found supporting the co-evolution of science and technology but the evidence is not strong:

Growth in the number of patents in a technological domain and growth in the number of scientific papers in the scientific super-discipline that is the most highly cited by patents in the domain are strongly correlated for some domains but not for others;

When the growth of patents in a domain is exponential with time, the exponent for growth of papers in the most highly cited scientific categories has a Pearson correlation coefficient of 0.617 with the exponent for patent growth showing strong but not complete correlation particularly considering the domains where patent growth is non-exponential.

Prior research that identifies emerging patent clusters and independent prior research identifying emerging scientific topics show statistically significant but qualitatively weak inter-relationships between the clusters and topics. This work also offers evidence that patent cluster emergence can, but does not usually, precede the emergence of related scientific topics.

The lack of clear evidence for co-evolution is interpreted as resulting from the documented complex many-to-many relationship of science categories and technological domains and is not considered evidence against co-evolution.

## 1. Introduction

The intimate coupling of science and technology is well-known but has proven to be difficult to specify. Vannevar Bush (Bush, 1945)'s linear model of science leading to technology has been widely criticized (Kline and Rosenberg, 1986; Stokes, 1997; Edgerton, 2004) but nonetheless the model has been seen (for example, Balconi et al. (Balconi et al., 2010) and Rosenberg (Rosenberg, 1990)) as having important elements correct-somewhat autonomous but coupled trends in science are linked to important technological trends. However, this modest agreement recognizes that determining the direction of influence between science and technology, determining the dominant mechanisms for the interplay between them as well as determining the total social influences in such interactions are complex, unresolved issues. The major aim of this paper is to attempt to make some contribution to these issues by decomposing both science and technology further than attempted by previous research on this aspect of technological change.

Klevorik et al. (1995) and Narin et al. (1997) both established that the interaction of science and technology is mediated by what areas of science and technology is examined. Klevorik et al. (1995) used results from the Yale industry survey to establish that various industries have interactions with different scientific fields and thus potentially have different technological opportunities. The scientific fields included in the survey include five basic sciences (biology, chemistry, geology, math and physics) and six fields of applied science (agricultural science, applied math, computer science, materials science, medical science and metallurgy). Their results indicate some interactions that- at their relatively broad level of decomposition- were focused (for example the drug industry interacts primarily with only biology and chemistry) and others that were much broader (for example the electronic components industry with everything except biology). Narin et al. (1997) in their wide-ranging and often detailed study of the citation of scientific papers by patents note that "..there is throughout the linkage phenomenon a subject-specific couple between the technology and the science upon which it is building". However, the scientific fields they consider are even broader than Klevorik et al (the five categories they use are clinical and biomedical research; chemistry; physics; engineering and technology; other fields). The current study relies on more recent research on the relatedness of scientific fields (Klavans and Boyack, 2009; Rafols et al., 2010) which utilized citations among scientific papers to classify journals into more than 150 distinct scientific categories. Rafols et al. (2010) have also used citation pattern analysis to show affiliations among these categories that support the definition of 16 super-disciplines which we also report here: the super-disciplines are narrower than but closer in granularity to the earlier work by Klevorik et al. (1995) and Narin et al. (1997).

The research reported here particularly pursues decomposition of technology further and less ambiguously than the prior work. Klevorik et al. (1995) decompose industry into 130 mainly four digit SIC level lines of business : industries are known to have only a rough correlation to technological differentiation (Kortum and Putnam, 1997; Schmoch et al., 2003; Lybbert and Zolas, 2014). Narin et al. (1997) roughly differentiate technologies based on patent classes at high levels of abstraction. A major technology decompostion applied throughout this paper utilizes the classification overlap method (COM) developed by Benson and Magee (Benson and Magee, 2013) to find the relevant and complete set of patents that represent a particular technological domain[1]. The technique was more broadly applied in a second paper (Benson and Magee (2015A) and later sets of patents that represent defined domains were found by Benson and Magee (2016) and Guo et al. (2016). This prior work establishes patent sets for 44 broadly arrayed technological domains which are examined in the current work.

The current paper also reports results from an even finer-grained set of "emerging patent clusters" as developed by Breitzman and Thomas (2015). Whereas our domains each contain several hundred to thousands of patents, the emerging clusters typically contain only ~8 patents. The clusters are identified by an algorithm that uses advanced patent citation techniques to find small numbers of patents in close to real time that are linked to suddenly highly cited patents. We also report results from a more fine-grained set of scientific papers than represented by the categories that are our basic unit of analysis for science. This is the "emerging topics in science and technology" identified by Small et al (2014). These emerging topics are identified by combining two large scale models of the scientific literature, one based on direct citation, and the other based on co-citation using a difference function that rewards clusters of scientific papers that are new and growing rapidly. The "emerging topics" are on average about 200 x smaller than the scientific categories we use elsewhere in this paper. Both the Breitzman and Thomas "emerging patent clusters" and the Small et al "emerging topics" are identified for the same years (2007-2010) thus affording a more fine-grained and near real-time examination of the science –technology relationship.

---

[1] In this paper, we utilize the definition of a technological domain used by Magee et al. (2016) – "artifacts that fulfill a specific generic function utilizing a particular, recognizable body of technical knowledge".

More rigorous and finer-grained decomposition cannot be expected to resolve the full complexity of the science and technology relationship given the instutional scale of each construct (science and technology) and given the large number of network, evolution with time, economic, social and institutional factors that have been identified as important or potentially important. Nonetheless, the ability to look for co-evolution and specific mechanisms for interaction is only possible if the correct specific category of science and the correct related technological domain are the focus. Such problems require not just holistic integration but appropriate decomposition as well.

The tool we use for finding linkages among our decomposed seience and technological constructs is a version of one used originally by Carpenter et al. (1980) and later by others ( Narin et al., 1997; Tijssen, 2001; Acosta and Coronado, 2003), namely examination of the citations by patents to the non patent literature. As will be detailed in section 3, we focus on references to Journals recognized by the Web of Science (WoS) and eliminate references to less scientific sources. As shown in the research by Meyer (2000), even these citations are not usually evidence that the new scientific knowledge in the paper is critical to the patent that cites it (in fact in our 44 domains containing 605,212 patents citing 395,338 scientific papers, we know of only one case that cleanly represents the direct mode often imagined where new science leads more or less directly to new technology -the spin-valve sensor for magnetic information storage based upon a discovery that later won a Nobel prize (Cros et al., 2009; Bajorek, 2014)[2]. However, the meaning we take for citations of scientific papers by patents- the scientific category of the knowledge has relevance to patents and artifacts in the technological domain of the patent citing the paper- appears to be consistent with Meyer's analysis.

The paper is organized as follows. In Section2, we describe the theoretical framework and develop hypotheses for later testing. Section 3 describes the data and the empirical strategy. Section 4 presents and discusses the results of our analyses. Section 5 concludes.

**2. Science and technology inter-relationship**

---

[2] There may well be others but possibly only a handful.

## 2.1. Scientic and technological decomposition –science categories, science super disciplines and technology domains

A major aspect of technological development is understood to be ongoing change in the knowledge underlying the technological domain of interest. Although this factor has been noted by many, a particularly well-known formulation due to Dosi (Dosi, 1982) discusses trajectories and paradigms. Dosi's discussion largely focuses on technological aspects (scale, function and history) as being the agents of change but he also notes the increasing role of scientific inputs in the innovative process. Others are more explicit about a direct role of science in technological change; for example, Arthur (2007) defines a technology as a means to fulfill a purpose by exploiting some effect and emphasizes new effects-presumably discovered by scientific work- in his discussion. Magee et al. (2016) define a technological domain as "A technologically differentiated field consisting of artifacts that fulfill a specific generic function utilizing a particular, recognizable body of technical knowledge". This definition is also utilized here as it, distinct from the others, allows us to find appropriate patent sets (Benson and Magee, 2015A). Finding the relevant patent set for a domain is necessary in order to identify the science cited by a domain –a necessary step in addressing our research questions that compare various domains and scientific categories based upon citations by the patents that are domain specific.

We utilize the comprehensive and multidisciplinary JCR journal analysis and evaluation reports for our decomposition of scientific publications by fields. In this system, all the journals included in WoS are analyzed. The ISI (Onex Corporation & Baring Private Equity Asia) assigns journals into different scientific fields based on journal-to-journal citation patterns and editorial judgment. Rafols et al. (2010) created a global science map based on citing similarities among different scientific fields. The data were harvested from the JCR of 2007, comprising 172 scientific categories and 55 social scientific categories. The 172 scientific categories were distributed into 16 super-disciplines. We adopt their classifications in our study, and research the scientific relatedness of different technological domains from both scientific categories and super-disciplines perspectives.

## 2.2. Science and technology evolution

Science and technology have generally been viewed as interacting, but somewhat autonomous systems. Price (1965), as well as Toynbee before him, regarded science and technology as "dancing partners" with no consistent leader and as different but reciprocal constructs (Rip, 1992). Debate about the direction of knowledge flow is central in papers such as Rosenberg (1982) who asked "how exogenous is science", indicating how technology often leads and precedes science and Klevorik et al. (1995) who argued for a role of universities and science as an important source of "technological opportunities" for industrial innovation. Arthur (2007) described phenomena/effects and principles that connect these phenomena to purposes. His analysis argues that radical invention often uses new principles to achieve the purpose while incremental ones use generic or existing principles. Similarly, Basnet and Magee (2016) also noted the importance of new basic oprating principles to the ongoing process of technological change which encompasses both science and technology.

Murray (2004) showed that academic inventors make contributions to firms at two distinctive levels-human capital and social capital. By identifying and analyzing a patent-paper pair and its networks, Murray (2002) explored the co-evolution of science and technology in an emerging area of biomedical science. Her study highlighted the considerable overlap existing between scientific and technological social networks.

*2.3. Correlation of decomposed science and technology*

Despite the differences in emphasis in different papers, the apparent consensus is given in the paper by Nelson and Rosenberg (1993) who stated clearly the intertwining of science and technology as a key characteristic of national systems of innovation (NIS). They recapitulate the complex interactions between these two dimensions highlighting that science is both "a leader and follower" of technological progress. A very similar conclusion is arrived at by Balconi et al. (2010). In this paper, we look for further specific and objective evidence for this point of view. Murray (2002)'s study showed that important aspects of the interaction are not revealed in either patents or scientific publications. Thus, we do not propose to clarify all aspects of the relationship between science and technology but instead to pursue the objective information opened up by use of technological domains as the appropriate structure for technology and scientific categories as the appropriate structure for science. It is anticipated that the addition of such objective information to the rich qualitative understanding available (Nelson and Rosenberg, 1993; Mokyr, 2016) will aid in understanding this complex interaction.

Despite the complexities and detailed differences, all theoretical and empirical work argues for connections between scientific and technological knowledge. Most of these treatments also explicitly or sometimes only implicitly recognize that the scientific knowledge associated with technology is domain dependent. We extend this implication to arrive at the first hypothesis that we test in this paper:

**H1A.** Relatedness among technological domains determined from similarity in citing specific scientific categories is correlated with relatedness among the same technological domains found only through patents citing patents in other domains.

An alternative possibility is that despite the known differences in the kind of science used in different technological domains, the scientific structure and the technological structure (or distance between fields) are uncorrelated. This is possible using Arthur's or our formulation (the technology structure may only relate to the function or purpose of the technology and not the underlying knowledge). Thus, our hypothesis states that (at least some of) the structure in technology is due to the structure of the knowledge in the related science.

Although there is little prior specific literature on the issue of breadth of links among SC and domains, discussion by Rosenberg (1979 and 1987) in several papers points to more diffused or broader impact and utilization of scientific inputs rather than singular scientific categories (or even fields or super-disciplines) aligning with singular technological domains. Thus our second hypothesis posits some limits to the correlation of relatedness in scientific and technological knowledge. In particular, we look for breadth of impact of scientific categories across a range of technological domains and for whether a specific technological domain is dependent upon the science in only one scientific category or even in only one super-discipline.

**H1B.** Specific Technological domains generally utilize scientific inputs from multiple super-disciplines and specific scientific categories contribute inputs to multiple technological domains.

*2.4. Dynamics of science and technology*

Beyond these hypothesized structural links to related scientific knowledge, technology and science are widely recognized as time dependent (Nelson and Rosenberg, 1993; Klevorik et al, 1995; Hunt, 2010; Magee, 2012; Benson and Magee, 2015B; Magee et al, 2016). Following this line of thinking and the concept of mutual causes, the dynamics of technological change are linked to changes in the cited scientific fields over time. Thus hypothesis 2A is:

**H2A.** Technological structure and scientific structure co-evolve.

This hypothesis relies upon the widely held belief that scientific structure and technological structure both change over time. Fundamentally, the hypothesis states that the structural changes in each side of the science and technology pairing are influenced by the changes in the other side.

As was mentioned in section 1, emerging technological clusters and emerging scientific topics were recently identified by researchers (Small et al, 2014; Breitzman and Thomas, 2015). If science and technology affect each others evolution, then the most important new technological fields might be expected to reflect (through citations) strongly the emerging scientific topics at a given time. Thus, our final hypothesis is as follows:

**H2B.** Emerging technological clusters and emerging scientific topics from 2007-2010 are closely related.

**3. Empirical strategy**

One of the most efficient and objective methods of evaluating research and innovation performance is through scientometric indicators. Publications (papers) that report theoretical and empirical research findings are the main channel for documentation and dissemination of scientific findings to further the development of science (Grupp, 1996; Schmoch, 1997). The methodology for testing the four hypotheses generated from prior research concerned with the interaction of science and technology from large citation databases used in this research was based first on the use of sets of patents that represent real-world technological domains very well. As mentioned above, we take advantage of a previously developed patent search technique [the COM, Benson and Magee (2013, 2015A)] to define 44 diverse technological domains along with highly relevant and complete patent sets representing each of these domains. The scientific paper citations generated by the 44 diverse technological domains patent sets and by the overall US Patent System (USPS) were retrieved as described in section 3.2.1. Testing of the first hypothesis (1A) is by analysis of the complex but differentiated scientific citation patterns for technological domains. The analysis of these patterns proceeds by calculation of vectors for science as determined by the 176 scientific categories defined for publications by the JCR (Journal Citation Reports) and the 16 super-disciplines defined by Rafols et al. (2010). The next step in the analysis is to use the science vectors to place the 44 domains on a technological map, and then compare it with the technology maps based only on domain cross citing of the patent data. Testing of the hypothesis H1B is by analysis of breadth of super-disciplines cited by specific domains and the breadth of domains citing specific scientific categories. Testing hypothesis H2A is performed by comparing time dependencies of patent and publication numbers in linked domain/category pairs. H2B is tested by studying the relative intensity of citations, the time relationship of citations among the emerging patent clusters and the papers in the emerging scientific topics and the diversity of linkage format between emerging patent clusters (Breitzman and Thomas, 2015) and emerging scientific topics (Small et al, 2014) from 2007 to 2010.

*3.1. Dataset*

*3.1.1. 44 technological domains patent data*

The initial 28 of the 44 domains patent database was developed by Benson and Magee (2013, 2015A). The extension to 44 domains was deliberately done to broaden the original 28 domains into software and biomedical domains and was reported by Benson and Magee (2015B) and Guo, Park and Magee (2016). The final database covers a broad range of technologies, including a total of 605212 patents. In order to find the relevant and complete set of patents that represent a particular technological domain, Benson and Magee developed a relatively simple, objective and repeatable method called the COM. The methodology consists of using keyword search in US patents since 1976 to locate the most representative international and US patent classes and then determining the overlap of those classes to arrive at the final set of patents. Fig. 1 shows a schematic of the process (Benson and Magee, 2013). The initiating pre-search was done using the patent search tool PatSnap, which searched all U. S. Patents from 1976 to 2013. Fig. 2 shows the size and relevancy of patent sets for all 44 technological domains. The overall size of the patent sets ranges from 154 (Flywheel Energy Storage) to 149491 (Integrated Circuit Processors) and typically 80-90% of the patents are found relevant to the domain by reading of samples (Benson and Magee, 2013; Benson and Magee, 2015A).

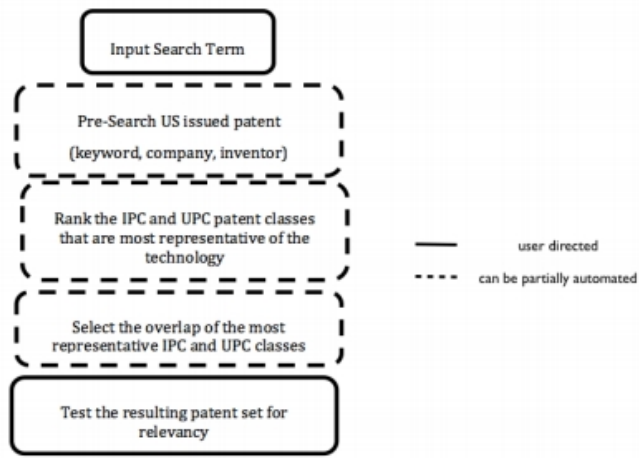

**Fig. 1.** Process flow of the COM (Benson and Magee, 2013). Most of the method can be automated via a computer, with only the selection of the search query and the testing of the final results left to the user. Step 1 is pre-searching US issued patent titles and abstracts for the search terms. Using the patent search tool PatSnap is an easier way. The input to the COM is simply a set of search terms that can be entered into a text box; Step 2 is ranking the IPC and UPC patent classes that are most representative of the technology; Step 3 is selecting the overlap of the most representative IPC class and UPC class which takes advantage of the different sense in which these classification systems were used by the highly experienced patent examiners.

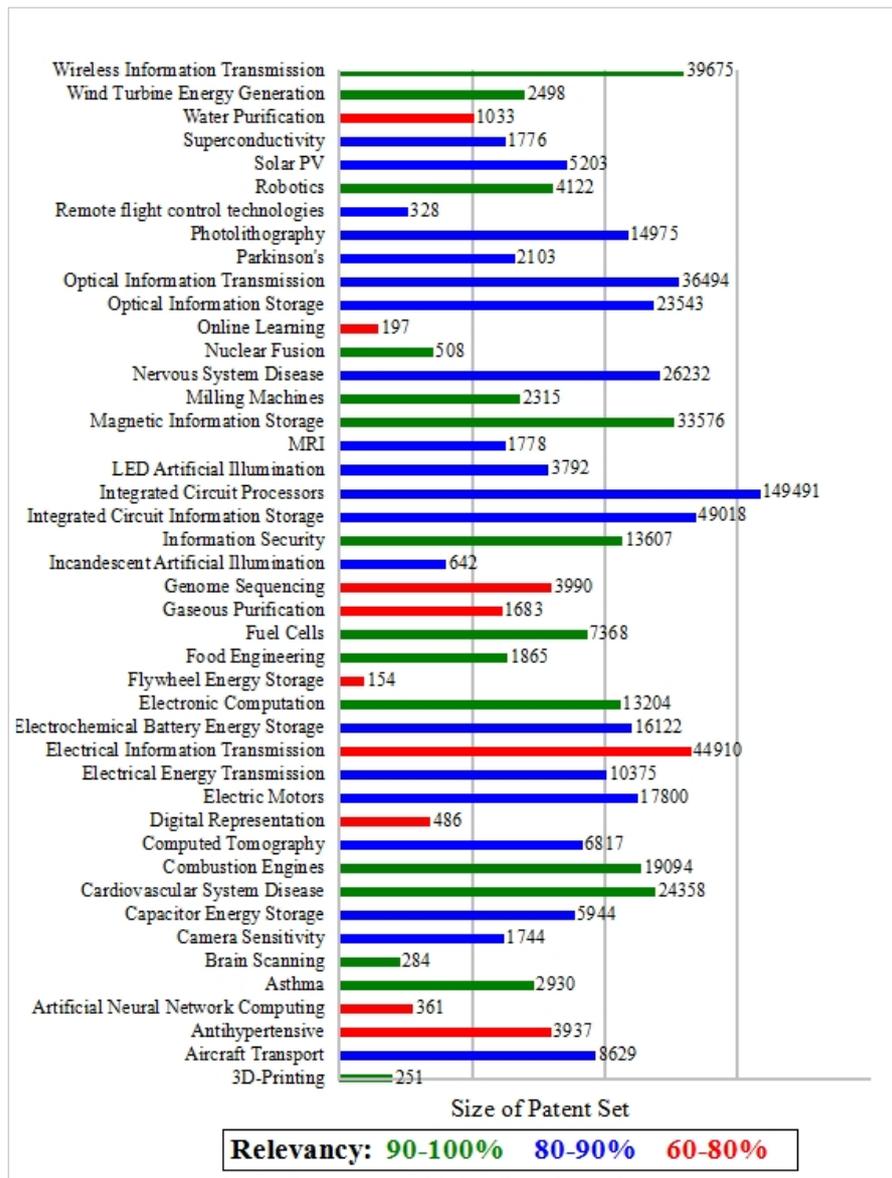

**Fig. 2.** Size (on log scale) and relevancy (in color) of patent sets for all 44 technological domains. This figure shows the wide range of technologies examined and the diverse size of technological domains (from 154 to 149491).

*3.1.2. Papers in emerging topics and patents in emerging clusters*

The data for the emerging scientific topics in 2007-2010 were provided by Small et al. (2014). Based on over 17 million articles in the Scopus database (1996-2010), Small et al. (2014) have created two large-scale models based on co-citation and direct citation respectively. The co-citation model was created using a multi-step process. First, clusters of cited papers are created for each separate year in the citation database. Second, current papers from the annual slice are assigned to the clusters of cited references. Finally, clusters from adjacent years are linked using shared reference papers into cluster strings. Creation of the direct citation model is much simpler. Citation links between articles are used to create clusters of articles using the full set of Scopus articles in a single clustering process. They then combined the two models to calculate *Emergence Potential* to identify the top 25 emergent scientific topics, which are new and growing rapidly for each year 2007 through 2010. Since a topic can last for more than one year, 71 emerging topics containing 31843 scientific papers (1996-2010) were identified overall.

The emerging patent clusters in 2007-2010 are due to Breitzman and Thomas (2015) who identified 6593 different emerging clusters between 2007 and 2010 based on all the US patents between 1980 and 2011. First, they identified the 'hot' patents that were cited highly in the most recent time period and represent a high percentage of total citations (these are often older patents). Second, they identified and clustered the next generation patents citing the hot patents for a subject year, and ranked the next generation clusters based on characteristics (for example, public sector proportion and science index) of the patents contained within them. Specifically, the next generation patents for year T are the patents from year T and year T-1. And third and last, 6593 emerging clusters containing 55650 different patents were defined as the 'emerging clusters' for the time period 2007-2010. Each cluster contains an average of ~8.4 patents that were published between 2006 and 2010, and one patent can belong to more than one cluster.

*3.2. Scientific vectors associated with the 44 domains*

*3.2.1. Vectors from 176 scientific categories*

The scientific vectors for technological domains are determined by the 176 scientific categories defined for publications by the JCR of 2013. The JCR are comprehensive and multidisciplinary journal analysis and evaluation reports, published by the Institute for Scientific Information annually. All the journals included in WoS are analyzed. The ISI assigns journals into different scientific fields based on journal-to-journal citation patterns and editorial judgment.

In this study, the raw data from each patent set contain citations to a variety of types of non-patent-literatures (NPL), which we separate into two kinds of references: papers from scientific journals (i.e. scientific papers) and others. Only the scientific papers included in the WoS were used in this study.

To find the scientific categories cited by patents, we download all the 176 scientific categories presented in JCR with their corresponding journal titles. The method we use recognizes the diversified formats of the scientific references of patents: each of the journal titles contains 3 formats: full journal title (ex: AAPS JOURNAL), abbreviated journal title (ex: AAPS J) and abbreviated journal title with dot (ex: AAPS.J). There are more than 8500 journals in total and one journal may belong to more than one scientific category (which makes the scientific vector different from citation's fraction as discussed later). For example, the journal *BIOMATERIALS* belongs to 'Engineering, Biomedical' and also to 'Materials Science, Biomaterials'.

After this, we use the string matching method in Matlab to process the scientific papers cited by patents. This method retrieves for each paper cited in a patent the corresponding journals as included in WoS. The JCR lists all of the more than 8500

Journals they cover in 176 scientific categories and we use this listing to determine the appropriate scientific categories for each citation.

Note: to get a more accurate result, when we use the string matching method, we removed ∼150 short-title journals like 'AGE', which may appear in the references very possibly not as a journal name but as part of the title since we found incorrect entries when we did not do this and their elimination did not make large differences in totals.

We then calculate the fractions of citations to each of the 176 scientific categories, take the fraction of one cited scientific category as the component of a vector, and we thus obtain the vectors based upon the 176 scientific category space.

*3.2.2. Vectors from 16 scientific super-disciplines*

For super-disciplines, we adopt the classification method proposed by Rafols et al. (2010). They obtained data from the CD-ROM version of the JCR of the Science Citation Index (SCI) and the Social Science Citations Index (SSCI) of 2007, comprising 221 categories. The 221 categories contained two classes: 172 scientific categories and 55 social scientific categories (6 of the categories belong to both classes). These data were used to generate a matrix of citing categories to cited categories with a total of 60,947,519 instances of citations among all scientific and social scientific categories. Salton's cosine was used for normalization in the citing direction. Then they use SPSS for factor analysis of the matrix and obtained 18 super-disciplines for 221 categories. The 172 scientific categories were agglomerated into 16 super-disciplines thereof to which they attributed the names shown in Table S1.

In contrast to the JCR for 2007, the number of scientific categories in JCR for 2013 which we used increased to 176 and the number of social scientific categories (not included in our study because they were rarely cited by technical patents) increased to 56 (as with the JCR of 2007, there are still the same 6 categories belonging to both classes). These increases occurred because new journals were added into the WoS after 2007 (the number of journals indexed by SCI increased from 6426 in 2007 to 8539 in 2013[3]), and a very small part of journals' scientific categories were changed (most of them were just added into one of the new scientific categories but also belong to their original categories). The impacts of these small differences were found to be insignificant: the 4 new scientific categories (audiology & speech-language pathology, cell & tissue engineering, logic and primary health care) are very rarely cited by technical patents (see the Null category in the Table S1). Thus, the names of the 176 categories given by JCR and their correspondence within the 16 super-disciplines are shown in the Table S1. The scientific categories are not evenly distributed among the super-disciplines; for example, there is only one scientific category (engineering, industrial) belonging to the super-discipline Business and Management, while there are 20 belonging to the super-discipline Engineering.

Based on the distribution of the 176 scientific categories in the patent set, the fraction of these domain citations for each of the 16 super-disciplines is simply

---

[3] The data obtained from http://wokinfo.com/products_tools/analytical/jcr/

arrived at by addition and thus the required super-discipline vectors are determined.

*3.3. Technological correlation of the 44 domains*

We study the technological correlation among 44 domains based on the domain cross citing of patent data. For a specific domain, we calculate the fraction of patents in each of the other 43 domains that are cited by all patents in this domain. We then take the fraction of one cited technological domain as the component of a vector, and we thus obtain the vectors based upon the 44 technological domain space.

**4. Results and discussion**

*4.1. The correlation between scientific relatedness and technological relatedness among technological domains*

In this section, in order to test Hypothesis H1A, we generate and examine two overlay maps of 44 technological domains based on their scientific correlation matrix (similarity to cited scientific categories determines distance) and technological correlation matrix (similarity to domains of cited patents determines distance) respectively. We also calculate the Pearson correlation coefficient of the two matrixes to quantitatively determine the correspondence of technological and scientific distances among the 44 domains. Then, to test Hypothesis 1B, we study the breadth of interaction from domains and scientific fields by examining the number of scientific fields cited by specific domains and how many domains cite specific categories.

*4.1.1. Scientific correlation matrix of 44 technological domains*
The scientific correlation matrix was generated based on the 44 technological domains' scientific vectors. The full vectors for all 44 domains are shown in Table S2, and the top 10 most cited scientific categories for each of the 44 domains are shown in Table S3.

After calculating the Pearson correlation coefficient of the 44 technological domains' vectors, we obtain the scientific correlation matrix for the 44 technological domains. The full symmetric matrix for all 44 domains is available in Table S4. Based on the 44 technological domains' 176 scientific vectors, one can easily get their scientific vectors in the 16 super-discipline space (the full vectors for 44 domains based on 16 super-disciplines and the corresponding correlation matrix are available in Table S5).

*4.1.2. Maps*
*4.1.2.1. Map based on scientific correlation matrix*
We generate the input for a map by using the Pearson correlation matrix of technological domains based on their scientific vectors. Fig. 3 shows a map of 44 domains based on the correlation matrix at the 176 vectors level (Table S4).

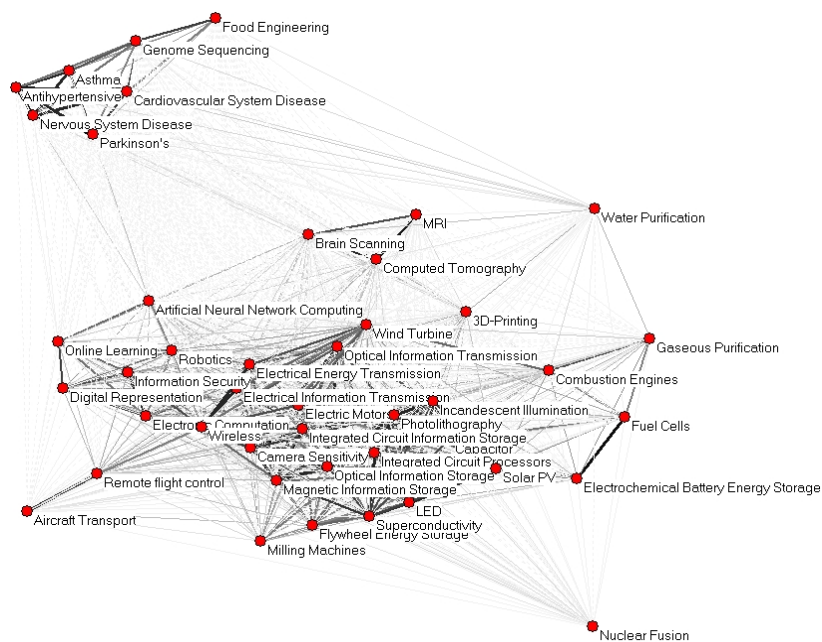

**Fig. 3.** The technological map of 44 technological domains based on 176 scientific vectors space. Pajek used for the visualization.

The map shows 44 domains in a Kamada-Kawai layout (using Pajek) that provides a representation of distances between different technological domains. Fig. 3 shows that 7 technological domains in the upper left corner (food engineering, genome sequencing and so on) have quite strong relationships with each other- that is all 7 domains cite papers in similar scientific categories, while distant from ( less related to) other domains because of low similarity in scientific citation categories. Some domains, such as aircraft transport, water purification and particularly nuclear fusion, are not strongly connected scientifically to any other technological domains. Other technological domains, such as 1) electrochemical battery energy storage and fuel cells, 2) superconductivity and LEDs and 3) the trio of brain scanning, MRI and computed tomography and others toward the center of this figure, have close relationships with each other but are not so distant from various other domains.

*4.1.2.2. Map based on technological correlation matrix*

Different from the Fig. 3, Fig. 4 was generated based on the cosine-similarity correlation matrix based on the inter-domain citing by patents (the full technological correlation matrix of 44 domains is available in Table S6).

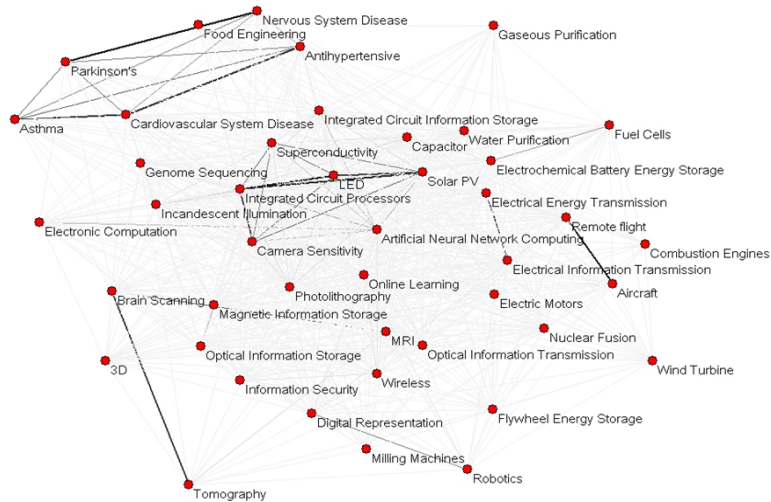

**Fig. 4.** The technological map of 44 technological domains based on the inter-domain citing of patents. Pajek used for the visualization.

Comparing Fig. 4 with Fig. 3, the 7 strongly-related technological domains in the upper left corner are exactly the same in the two figures but somewhat closer to the other domains in Fig. 4 than Fig. 3. Moreover, fuel cells and batteries, as well as superconductivity and LEDs are close in both maps. There are other similarities but there are also some clear differences. One example of a difference is that nuclear fusion and aircraft transport (and its nearest neighbor remote flight control) are much closer to electric motors and some other technological domains in Fig. 4 than in Fig. 3. Another clear difference in the opposite sense is that robotics and electrical energy transmission are much closer together in Fig. 3 than they are in Fig. 4 indicating they reference papers in similar scientific topics more than they reference patents in similar domains. A third more general difference is that the scientific knowledge separation depicted in Fig. 3 between energy technologies and information technologies is much less than the technological separation for energy and information technologies in Fig. 4 (for example, compare the surroundings of flywheel energy storage, wind turbines and electric motors in the two charts). The qualitative similarities give support to hypothesis one showing that the structural relationships in science of the fields cited by technological domains (Fig. 3) is reflected when one maps the same domains just based upon the which domains cite patents from each other. Moreover, the wider separation of energy and information technological domains in the technological map (Fig. 4) and other detailed differences between the figures suggests that relatedness is promoted by technological factors beyond similarity in fundamental scientific knowledge characteristics.

*4.1.3. Quantitative correlation of science and technology matrix*

To quantitatively characterize the correlation between the two matrices based on the scientific vectors and domain cross citing respectively, we calculate the Pearson correlation coefficient of these two matrixes (by using Matlab), the result is 0.564, which like the qualitative map shows a relatively close but quite far from perfect relationship between the scientific knowledge similarity and the patent citation similarity of the technological domains. These results support but also qualitatively and quantitatively refine hypothesis H1A.

*4.1.4. Breadth of interaction from singular domains and scientific categories*

In regard to information about hypothesis H1B concerning the breadth of use of science by a singular domain, we note that in 2010 that domains cited on average 8.3 super-disciplines and 36.3 categories. This clearly demonstrates multiple scientific categories for domain citation of science (supporting H1B). Another measure of the breadth is that the top 3 SC cited by a domain account for about ½ (~0.494) of the total citations demonstrating a degree of concentration. However, the average domain does not cite the majority of its scientific references from a single super-discipline which overall suggests wider science input to a domain than is often assumed.

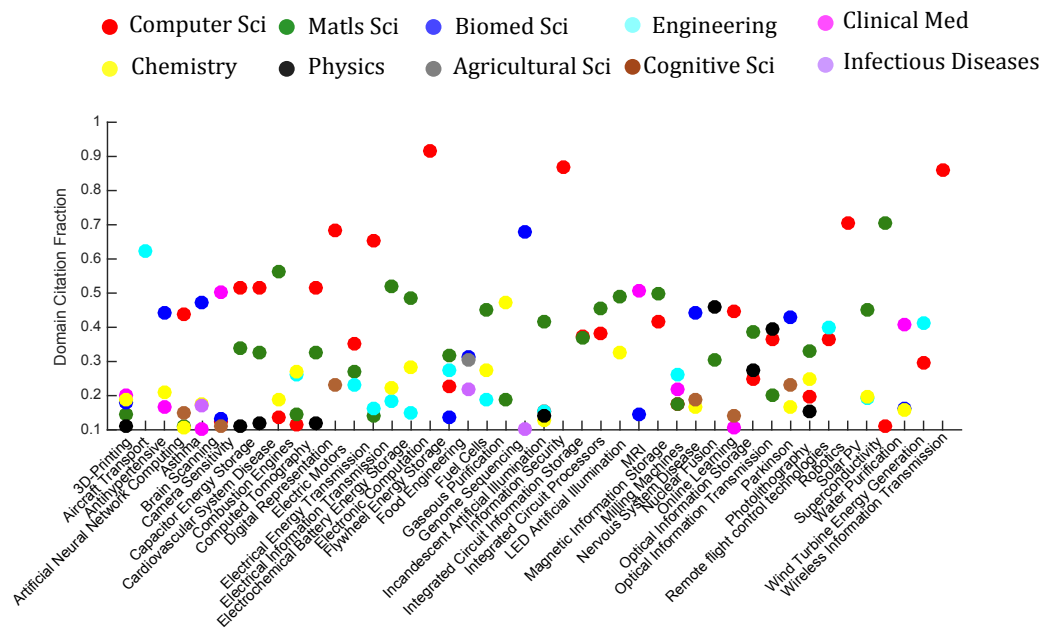

**Fig. 5.** 44 domains and scientific citations in the highest cited of the 16 super-disciplines. For each technological domain, all scientific citations by patents from 1976-2013 were counted and classified in the 16 super-disciplines with the fraction of these domain citations for each of the super-disciplines given on the ordinate. The super-disciplines receiving 0.1 or more of the citations in a domain are plotted.

The 44 domains and their distributions of scientific citations among the 16 super-disciplines are given in Fig. 5 for the 10 most highly cited super-disciplines. The computer science super-discipline receives approximately 90% of the citations in the electronic computation domain, the information security domain and the wireless domain: physics receives the most citations in the optical information transmission domain (~40%) but computer science (>35%) and materials science (~20%) also receive significant numbers of citations in this domain. The materials science super-discipline is strongly cited in most of these 44 domains and is the most highly cited in ~1/3 of these domains. Note that citations by patents to scientific papers in Journals that are in the materials science super-discipline have increased overall but the fraction of such citations to all citations has decreased in the last 20 years. This is due to "explosive" growth of citations to software and biomedical super-disciplines during the same period.

The most widely cited scientific category in the 44 domains is engineering, electrical & electronic, which is cited by 33 out of the 44 domains showing input from some scientific categories to many domains. Moreover, the average SC contributes to ~10 different domains. These findings also generally support hypothesis H1B.

*4.1.5. The distribution of cited scientific categories*

Fig. 6A shows the distribution of scientific citations among the 176 categories for all patents in the 44 domains in 2010. They are plotted on a log scale (fraction of cites) and ranked based upon the cited fractions in descending order. The result is nearly a downward sloping straight line from left to right except for ~ the highest 10 ranked categories and the lowest ranked 20 categories. This result is closely similar to the distribution of scientific citations in *all* the US patents in 2010 (Fig. 6B). The similarity includes the slope and the rough magnitudes of the citation fractions at various rankings.

A.

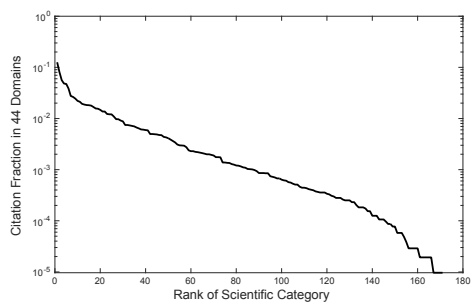

B.

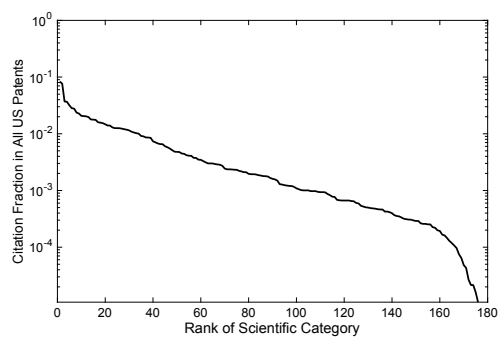

**Fig. 6.** Distribution of the cited scientific categories. (A.) by patents in 44 domains published in 2010; (B.) by all US patents published in 2010.

An additional important point to note is that the top 5 most cited scientific categories for 44 technological domains (*engineering electrical & electronic*, *applied physics*, *pharmacology & pharmacy*, *biochemistry & molecular biology* and *optics*) are also in the list of top 10 categories for all US patents in 2010 (shown in Table 1). These results indicate that, our 44 domains- as intended- are a fair representation of all US patents but the modest switching in ranking between the two lists indicates that the 44 domains are not a perfect sample drawn from the overall patent database.

**Table 1**

Top 10 cited scientific categories by patents of 44 domains and all US patents.

| Rank | Scientific Category Cited by 44 Domains | Cited Fraction | Scientific Category Cited by All US Patents | Cited Fraction |
|---|---|---|---|---|
| 1 | ENGINEERING, ELECTRICAL & ELECTRONIC | 0.126 | BIOCHEMISTRY & MOLECULAR BIOLOGY | 0.126 |
| 2 | PHYSICS, APPLIED | 0.080 | ENGINEERING, ELECTRICAL & ELECTRONIC | 0.091 |
| 3 | PHARMACOLOGY & PHARMACY | 0.056 | CELL BIOLOGY | 0.056 |
| 4 | BIOCHEMISTRY & MOLECULAR BIOLOGY | 0.048 | PHYSICS, APPLIED | 0.045 |
| 5 | OPTICS | 0.047 | PHARMACOLOGY & PHARMACY | 0.029 |
| 6 | NEUROSCIENCES | 0.039 | OPTICS | 0.026 |
| 7 | MATERIALS SCIENCE, MULTIDISCIPLINARY | 0.028 | BIOTECHNOLOGY & APPLIED MICROBIOLOGY | 0.024 |
| 8 | TELECOMMUNICATIONS | 0.026 | CHEMISTRY, MULTIDISCIPLINARY | 0.019 |
| 9 | PSYCHIATRY | 0.025 | MEDICINE, RESEARCH & EXPERIMENTAL | 0.019 |
| 10 | PHYSICS, CONDENSED MATTER | 0.022 | TELECOMMUNICATIONS | 0.018 |

*4.2. The co-evolution of technological structure and scientific structure*

The results supporting hypotheses 1A and 1B show that new technologies tap into different knowledge bases-some of which are clearly based upon different knowledge in different scientific fields but that any given technological domain is related to multiple categories and any given scientific category relates to numerous technological domains. To test hypothesis 2A about co-evolving structures, we look at the time dependence and the domain dependence of the science links. Thus, we look for relationships between output of patents in domains and output of papers in the scientific category most highly cited by the same technological domains.

*4.2.1. Co-evolution of technological output and scientific output*

The growth of the entire corpus of literature in the database of WoS, and the growth of total number of patents in 44 domains are each exponential with time and the growth of the two is therefore not surprisingly correlated. The total scientific papers in WoS from 1980 to 2010 increased from 749426 in 1980 to 2098092 in 2010 (a factor of ~1.8), and fits an exponential curve with an exponent 0.033. The growth of total patents also fits an exponential but with a higher growth rate of 0.072. Since both are exponentials, the Pearson Correlation Coefficient between these two series (paper growth and patent growth) is fairly high at 0.877 but this cannot be interpreted as evidence of co-evolution since two independent but exponential processes would have high correlation. Therefore, some more stringent tests between growth of patents in a given domain with the growth of papers in strongly related scientific categories were pursued.

The correlation between growth of patents in each domain and the total scientific papers[4] in the super-discipline, which was most cited by that domain was tested. The result in Table 2 shows a more or less continuous distribution from quite high correlation and low p values to poor correlation with high p values overall suggesting weak –or at best modest- correlation in the patent output and the publication time series. Although the highly correlated cases appear to support co-evolution, the existence of domains where that does not hold suggests that on balance, these results offer only weak support for co-evolution.

**Table 2**
The correlation between growth of technological domain and the most cited scientific fields.

| Technological Domain | Most Cited Scientific Field | Pearson Correlation | P- Value |
|---|---|---|---|
| Optical Information Storage | Material Science | 0.932 | <0.01 |
| Wireless Information Transmission | Computer Science | 0.925 | <0.01 |
| Optical Information Transmission | Physics | 0.892 | <0.01 |
| Fuel Cells | Material Science | 0.88 | <0.01 |
| Integrated Circuit Information Storage | Computer Science | 0.871 | <0.01 |
| Information Security | Computer Science | 0.866 | <0.01 |
| Integrated Circuit Processors | Material Science | 0.864 | <0.01 |
| Brain Scanning | Clinical Medicine | 0.852 | <0.01 |
| Magnetic Information Storage | Material Science | 0.81 | <0.01 |
| Computed Tomography | Computer Science | 0.794 | <0.01 |
| Electrical information Transmission | Material Science | 0.794 | <0.01 |
| Nervous System Disease | Biomedical Science | 0.793 | <0.01 |
| Asthma | Biomedical Science | 0.79 | <0.01 |
| Electronic Computation | Computer Science | 0.789 | <0.01 |
| Online Learning | Computer Science | 0.789 | <0.01 |
| LED | Material Science | 0.781 | <0.01 |
| Parkinson's | Biomedical Science | 0.778 | <0.01 |
| Photolithography | Material Science | 0.772 | <0.01 |
| Cardiovascular System Disease | Material Science | 0.768 | <0.01 |
| Digital Representation | Material Science | 0.766 | <0.01 |
| Capacitor | Computer Science | 0.765 | <0.01 |
| Electrochemical Battery Energy Storage | Material Science | 0.738 | <0.01 |
| Remote flight control technologies | Engineering | 0.734 | <0.01 |

---

[4] Note that, in this section, the total scientific papers means that all the independent scientific papers indexed by the database of WoS in the super-discipline- not only the papers cited by the patents in the technological domain.

| | | | |
|---|---|---|---|
| Genome Sequencing | Biomedical Science | 0.724 | <0.01 |
| Electric Motors | Computer Science | 0.72 | <0.01 |
| Nuclear Fusion | Physics | 0.714 | <0.01 |
| Camera Sensitivity | Biomedical Science | 0.684 | <0.01 |
| Combustion Engines | Chemistry | 0.66 | <0.01 |
| Water Purification | Clinical Medicine | 0.657 | <0.01 |
| Artificial Neural Network Computing | Computer Science | 0.598 | <0.01 |
| MRI | Clinical Medicine | 0.596 | <0.01 |
| Wind Turbine Energy Generation | Engineering | 0.58 | 0.001 |
| Flywheel Energy Storage | Material Science | 0.558 | 0.001 |
| Food Engineering | Biomedical Science | 0.554 | 0.001 |
| Aircraft Transport | Engineering | 0.503 | 0.004 |
| Antihypertensive | Biomedical Science | 0.442 | 0.013 |
| 3D | Clinical Medicine | 0.325 | 0.075 |
| Milling Machines | Engineering | 0.309 | 0.09 |
| Solar PV | Material Science | 0.217 | 0.241 |
| Robotics | Computer Science | 0.206 | 0.267 |
| Gaseous Purification | Chemistry | 0.201 | 0.278 |
| Superconductivity | Material Science | 0.189 | 0.308 |
| Incandescent Artificial Illumination | Material Science | 0.079 | 0.673 |
| Electrical Energy Transmission | Computer Science | 0.035 | 0.854 |

If science relative to a domain and technological output from a domain co-evolve, one anticipates that the faster growing scientific fields are associated with faster growing technological domains. To test this idea, we looked for domains that fit an exponential well so we could estimate the rate of growth accurately. We found that patent growth in 19 of the 44 technological domains was strongly exponential, and except the domain of MRI, the growth of total scientific papers in the top scientific categories[5] cited by these domains was also exponential. Table 3 shows the growth exponents for the 18 domains and the top cited scientific categories.

**Table 3**
The growth exponents of 18 technological domains and the top cited scientific fields.

| Domain | Domain Growth Exponent | Scientific Categories Growth Exponent |
|---|---|---|
| Aircraft Transport | 0.025 | 0.047 |
| Camera Sensitivity | 0.054 | 0.048 |
| Capacitor | 0.044 | 0.043 |
| Cardiovascular System Disease | 0.063 | 0.023 |

---

[5] For a technological domain, the number of top scientific categories we choose due to the proportion of scientific citation of these categories is enough categories to get >70% of the total citations. For example, for the domain of Camera Sensitivity, we take the top 2 categories (*engineering, electrical & electronic* and *applied physics*) for further analysis. The proportion of each scientific category for all 44 domains is shown in Table S2.

| | | |
|---|---|---|
| Combusion Engines | 0.034 | 0.017 |
| Computed Tomography | 0.077 | 0.033 |
| Electric Motors | 0.042 | 0.043 |
| Electrochemical Battery Energy Storage | 0.038 | 0.034 |
| Electronic Computation | 0.108 | 0.053 |
| Integrated Circuit Information Storage | 0.086 | 0.045 |
| Integrated Circuit Processors | 0.101 | 0.047 |
| LED | 0.138 | 0.059 |
| Magnetic Information Storage | 0.042 | 0.033 |
| Nervous System Disease | 0.067 | 0.057 |
| Optical Information Storage | 0.07 | 0.048 |
| Optical Information Transmission | 0.079 | 0.038 |
| Photolithography | 0.121 | 0.07 |
| Wireless Information Transmission | 0.097 | 0.078 |

The Pearson correlation coefficient of these two exponent series is 0.617, which indicates a moderately strong relation between the growth of technology and the supporting science. This result is interpreted as moderate support for the co-evolution hypothesis but it is weakened by the fact that 26 of the 44 domains did not show good enough exponentials to be included in the test. As a summary, H2A receives only weak support from this research.

*4.2.2. Co-evolution of emerging patent clusters and emerging scientific topics*
*4.2.2.1. Overall statistical tests of the links*

Hypothesis H2B requires a strong link between the emerging patent clusters identified by Breitzman and Thomas and the emerging scientific topics identified by Small et al. This section tests for such links by first examining the overall citation pattern by patents of the emerging topic papers and then comparing this to the citations from patents in the patent clusters.

After matching the papers in emerging topics identified by Small et al. (2014) with the scientific citation by all US patents in 2006-2010, we found that, in the 71 emerging topics, 57 are cited by patents, and 14 are not cited by any patent.

According to WoS, there are 18,528,862 papers published in 1996-2010 in total, whereas we find that ~143,098 of them were cited by patents (2006-2010). Thus, the ratio of all papers cited by patents is ~0.0077 (143,098/18,528,862), and in the 31,843 papers in 71 emerging topics identified by Small et al. (2014), we find 389 different papers cited by all patents (2006-2010). Thus, the ratio of papers in emerging topics cited by patents is ~0.0122 (389/31,843), ~1.6X the random expectation ratio (0.0077). Chi-square tests (Table 4) show that the two ratios have significant difference (sig.<0.001). Thus, the papers from the emerging topics are more often cited by patents than is expected randomly and this result is statistically significant; however the ratios of these papers actually cited by patents is still small (0.0122) and the factor lifting the emerging topic papers (1.6) while statistically significant does not appear to us to be qualitatively impressive.

**Table 4**

Chi-Square test of papers cited by patents.

|  | Value | Exact Sig. |
|---|---|---|
| Pearson Chi-Square | 83.663 | <0.001 |

The 14 emerging scientific topics that are not cited by any patent are shown in Table 5. First, there are no clear qualitative reason these topics should not be cited in patents (exception being number 27 -unparticle physics). Second, the 14 scientific topics contain 4582 papers (~327 papers/topic), and the other 57 scientific topics that are cited by patents contain 27261 papers in total (~478 papers/topic). Thus, fewer papers in these 14 non-cited scientific topics might be part of the explanation for why they are not cited by the patents. Third, compared to the papers in the 14 non-cited scientific topics, the 389 papers in the 57 cited- by-patents scientific topics are cited much more by *other scientific papers*. For all the papers in the 71 emerging scientific topics, the average cited frequency by other scientific papers is ~130 (until November 4, 2016). While, the average cited frequency by other scientific papers is ~816 for the 389 papers (6.28 X the expectation frequency). This is strong evidence that the emerging papers cited by patents also recieve more citations from scientific papers than other papers.

**Table 5**
Non-cited 14 topics.

| Non-cited ID | Cluster | #Papers |
|---|---|---|
| 1 | Fe superconductor | 1649 |
| 9 | theories of gravity | 211 |
| 11 | graphene optoelectronic applications | 188 |
| 16 | breast cancer | 318 |
| 19 | bocavirus DNA in children | 351 |
| 25 | graphene transistors | 179 |
| 27 | Unparticle physics | 201 |
| 30 | obesity-associated gene FTO | 257 |
| 35 | crystal packing is stabilized by inter-molecular O - H, N and C - HO hydrogen bonds | 246 |
| 49 | Crystallography | 114 |
| 52 | risk of acute myocardial infarction (AMI) following AMI hospitalization or stent insertion | 129 |
| 53 | Metabolic Syndrome | 279 |
| 54 | preventing influenza virus transmission | 190 |
| 70 | cognitive radio networks | 270 |

*4.2.2.2. Time difference*

Co-evolution of science and technology (H2A) suggests mixed time leadership between technology and science over time and that is tested here for the emerging patent clusters and emerging scientific topics. In the 6543 emerging patent clusters identified by Breitzman and Thomas (2015), only 142 clusters cite papers from 35 emerging scientific topics (out of 71) identified by Small et al. (2014). Thus, there are 36 emerging topics that are not cited by any of these emerging clusters.

There are 526 citations/links from 142 emerging clusters to 35 emerging scientific topics. Based on the time when clusters and topics have been identified, we found that, there are 415 times the emerging clusters cite papers in the emerging topics after the date when the topics are identified (~0.79), also 62 times at the same time (~0.12) and 49 even earlier than the topic is identified (~0.09).

These results indicate that Small et al. (2014) are correct that many of their topics have technological significance. It also is apparent that technological implications are sometimes seen (evidenced by early patent citations) before the scientific activity has accelerated enough to be identified by even a technique as sensitive as that used by Small et al. (2014). In a larger sense, this change of leadership between technology and science is support for the idea of co-evolution as opposed to a simple linear idea that technological activity always follows prior scientific activity.

*4.2.2.3. Multiple links*

In the 142 patent clusters that cite emerging topics, 82 clusters have the following 3 kinds of multiple links.

First, 80 patent clusters contain more than one patent having citation links from the patent cluster to the same scientific topics. Thus, the ratio of clusters having multiple patents linking the cluster and topic is ~0.563 (80/142).

There are 142 clusters citing 35 topics for 526 times in total, thus, to calculate the random expectation of the number of clusters that have this kind of link, we used Monte Carlo simulation and ran 100,000 simulations. The result is that 23.7 clusters is the random expectation and therefore, the random expectation ratio of clusters having this kind of link is ~0.167(23.7/142). The Chi-square tests (Table 7, 1st kind) show that the two ratios have significant difference (sig.<0.001).

There are only 6 patent clusters citing papers from more than one scientific topic: all six cases are shown in Table 6. The table shows that the topics cited in several of the cases are quite similar (for example, clusters 2, 3 and 5 all cite topics about *wireless*, and the two topics cited by cluster 4 are both about *graphene*). Thus, only cases 1 and 6 give possible evidence for clusters citing non-similar scientific fields. This is an indication that the clusters are more focused than domains and this is expected given the relatively small size of the clusters.

**Table 6**
Clusters citing papers from more than one topic.

| Cluster ID | Cluster | #Patents cite emerging topics | #Emerging topics | Emerging topics |
|---|---|---|---|---|
| 1 | 201000088 | 72 | 2 | transformation optics |
| | | | | graphene under strain; graphene mechanics |
| 2 | 201000461 | 12 | 2 | MAC protocols for applications of wireless sensor networks |
| | | | | wireless sensor networks |
| 3 | 201000479 | 16 | 2 | MAC protocols for applications of wireless sensor networks |
| | | | | wireless sensor networks |
| 4 | 201001040 | 5 | 2 | bilayer graphene nanoribbons with armchair edges |
| | | | | epitaxial grapheme |
| 5 | 201001216 | 19 | 3 | wireless sensor networks |
| | | | | MAC protocols for applications of wireless sensor networks |
| | | | | cognitive radio |
| 6 | 201001497 | 12 | 3 | human papillomavirus vaccine |
| | | | | some miRNAs can act as oncogenes or tumor suppressors |
| | | | | ultra high-throughput sequencing technologies |

Secondly, there are 24 clusters that have links from one patent in a cluster to several papers from the topic. We set the Monte Carlo simulation as 35 scientific topics cited by 259 different patents in 142 clusters for 526 times, and ran for 100,000 simulations. The expected number of clusters that have this kind of links is ~13.42. The Chi-tests (Table 7, 2nd kind) show that the two ratios have significant difference (sig.<0.05) but the significance is weaker than for the first case.

There is only 1 cluster containing a patent (US7653484) that cites two different topics. This patent cites two papers titled "*Performance Evaluation of Suvnet with Real-Time Traffic Data*" and "*Sequence-Based Localization in Wireless Sensor Networks*", which belong to two quite similar topics- *a mobile ad-hoc network (MANET)* and *wireless sensor networks* respectively. Thus, single patents (different than single domains) appear to largely have a more focused link to science.

Thirdly, in the 142 clusters that cite papers in emerging topics, 64 clusters have links from more than one patent in a cluster to an identical pair of papers in a scientific topic. There are 97 different papers in 35 topics cited by the 142 emerging clusters with 526 links among them, using the Monte Carlo simulation, we found that the random expectation is for 9.45 clusters to have this type of links specifically, we ran 100,000 simulations). The Chi square tests (Table 7, 3rd kind) show that the two ratios have significant difference (sig.<0.001).

Table 7 shows the Chi-Square tests results for these 3 kinds of multiple links between the actual frequencies and the random expectations demonstrating that cases 1 and 3 are particularly far from random.

**Table 7**
Chi-Square test of multiple links.

|  | Pearson Chi-Square Value | Exact Sig. |
|---|---|---|
| 1st kind | 47.576 | <0.001 |
| 2nd kind | 3.76 | 0.038 |
| 3rd kind | 55.775 | <0.001 |

The test results showing the significant differences between these three pairs of ratios (sig.<0.05), further manifests that different patents in the same cluster are more likely to cite the same topic; that one patent in a cluster tends to cite papers from the same topic; and that patents citing identical papers are very likely to be in the same patent cluster.

## 5. Discussion and concluding remarks

The research strategy followed in this paper has been to address the complex interaction of science and technology by focusing on more deeply decomposed descriptions of both science and technology than has previously been pursued. The major signal we have used for the interaction among the decomposed technology and the decomposed science is citations of scientific publications by patents. In section 2, we used past research to establish four hypotheses that were tested in the remainder of the paper. The first two of these hypotheses involved the structure or ontology of technology and its interaction with science. The latter two hypotheses involved the dynamics of the interaction between these two aspects of technological change. In this section, we restate the hypotheses and summarize our findings relative to each hypothesis while also more generally discussing the structure and dynamics of the science/technology interaction. We then note some limitations of the research and more broadly discuss the implications of the findings on our topic-interaction of science and technology in technological change.

**H1A.** Relatedness among technological domains determined from similarity in citing specific scientific categories is correlated with relatedness among the same technological domains found only through patents citing patents in other domains.

Our results confirm this hypothesis and show a Pearson correlation coefficient of 0.564 between the set of vectors based only upon the relationships of the *science (published scientific papers)* cited in the patents and the set of vectors based only upon the citations of the patents to patents in other domains- a *technological* structure descriptor. Although the correlation is quite strong, it is significantly less than 1 so underlying scientific knowledge is not the only differentiator among technologies. The qualitative comparison of maps drawn from the two measures of distance also confirms the hypothesis. Moreover, the qualitative examination of the maps supports the idea of a strong but not total correlation and that technological relatedness goes beyond fundamental scientific knowledge categories. This finding is consistent with the definition of a technological domain we use- that in addition to the scientific knowledge base, the second factor differentiating technological domains from one another is the function (purpose or basic utility) of the technological artifacts in the domain. Qualitatively, the two concepts –functional categories and knowledge categories- appear adequate to describe differences among technologies consistent with real world artifacts and performance of such artifacts (Benson and Magee, 2015B). No other categorization system for technologies appears able to accomplish this. In the research reported here, this decomposition of technology appears to allow one to adequately describe how scientific knowledge differences among domains affects the basic technological relatedness.

**H1B.** Specific Technological domains generally utilize scientific inputs from multiple super-disciplines and specific scientific categories contribute inputs to multiple technological domains

Our results confirm this hypothesis: we find that the average technological domain cited 8.3 super-disciplines and 36.3 scientific categories in 2010. Moreover, some scientific categories apparently contribute knowledge to a wide range of technological domains. The peak scientific category (engineering, electrical & electronic) is cited in 33 of our broadly chosen 44 technological domains and even the average SC is cited in ~ 10 domains. The concept of "spillover" is widely supported but is usually imagined to involve technological concepts developed in one technological field being applied in another field. The results reported here should remind all that "spillover" can be fostered or at least mediated by science. Our results show that this spillover is quite broadly operative across science and technology affecting multiple technological domains; narrow conceptual mechanisms for spillover appear inappropriate according to our findings.

The results supporting H1B show clear evidence that the science/ technology relationship involves many-to-many (technological domains and scientific categories) interactions and knowledge dependencies. While these results show broader distributions of science of a given kind throughout the technological enterprise, it is not at all random as the support for H1A indicates. The differences among technologies are partly due to differences in the scientific knowledge base upon which they rely despite the many-to-many underlying structure of the relationship. These findings are consistent with the qualitative concept of a complex relationship.

**H2A.** Technological structure and scientific structure co-evolve.

The finding that some emerging patent clusters (emerging technology) sometimes precedes the related emerging scientific topic is strong evidence for co-evolution but is weakened by the lack of strong linkage found between the emerging topics and emerging patent clusters (see H2B below). Moreover, the research reported here extensively examined the time dependence in the technological regime and the time dependence in the scientific regime but finds no highly reliable evidence to confirm that the two regimes co-evolve while importantly influencing one another. The evidence was mostly judged as weak with only one case of moderate support. We do not interpret this as evidence against the co-evolution hypothesis. It is more likely, in our opinion, that the interaction of science and technology is bi-causal or co-evolving but is also much broader than that between a given domain and its most closely associated science disciplines as support for H1B indicates. Our tests largely attempted to compare impacts and outputs among such domains and related disciplines and these tests would miss a broader co-evolutionary relationship.

**H2B.** Emerging technological clusters and emerging scientific topics from 2007-2010 are closely related.

Although statistically significant differences are found between citations to papers in the emerging scientific topics relative to those to scientific papers in general, the factors are not qualitatively impressive. Similarly, the citations to these papers by patents in the emerging clusters are also statistically greater than citations by random patents but again the statistical significance is not matched by qualitative strength. Nonetheless these results show that the clusters and topics are related but perhaps not as strongly as expected. Further interpretation of this result could again involve complexity but it could also simply be due to one or the other (or both) of these suggested methods not really identifying the technologically exciting topics despite both having resulted from a number of years of related studies. We simply do not yet know if a reliable methodology for identifying newly emerging important technologies exists and the work reported here indicates that we do not have *two* complementary, reliable methods. Thus, at our deepest attempt at technology and science decomposition, the results were less informative than we anticipated.

Our results document the increasing intensity of citations of scientific papers by patents but this finding cannot be considered as establishing the existence of ever closer links between science and technology: an unknown part-perhaps all- of the intensity increase is due to changes in patent citation practices over the past 40 years (Schmoch, 1993). In addition, our results indicate that there is no obvious lead/lag relationship between science and technology. Thus one must be careful not to conclude that science push vs. technological pull is established by the high citation of scientific papers by patents and the near zero citation of patents by scientific papers (Meyer, 2000): there are clear indications that the selection of scientific projects for funding and the selection of scientific disciplines for funding are influenced by technological need and opportunity (Small et al., 2014). On the other hand, finding a broad diversity of scientific disciplines contributing to different specific technological domains is a strong step in documenting the interaction of broad fields of science with the technological enterprise.

Our work and lots of other research addressing the science/technology interaction depicts this interaction as highly complex with many variables. A clear limitation of this research (and all other research involving the interaction of science and technology to our knowledge) is that the activities important in the interaction are far broader than the empirical foundation we used to address the problem. In particular, fundamental research, applied research, Pasteur's quadrant (Stokes, 1997) research - which involves intimate combining of fundamental and applied research, product development, manufacturing development, learning through manufacturing experience, customer research, profit analysis, technical social networks, "gatekeeping", spillover and many other activities are part of the science/technology spectrum. Our methods involved focus on publication which is a giant data source characterizing fundamental research (and partially applied research) and on patents which is another giant data source characterizing the output of applied research (and limited partial output from some other activities). The major study that looks at social networks (Murray, 2002) is a small number of case studies and has no quantitative grounding. Perhaps a viable methodology for large scale social network studies could be developed along the lines indicated by Pentland (2014). If this was done as broadly as the patent and publication system, important additional insights are possible. Moreover, research on documents other than patents and scientific publications is conceivable and the use of natural language processing looks promising if such documents become available. However, some of the most interesting documents for analysis (research proposals, strategy documents, product proposals, startup funding proposals, etc.) are likely to remain confidential and private. Thus, the current work has focused on the largest scale, accessible and relevant data sources for understanding the interaction of science and technology even though they are not sufficient for a full analysis of this complex interaction.

## Acknowledgments

The authors acknowledge the funding support for this work received from the SUTD-MIT International Design Center (SUTD is the funder, and the grant number is 6921538) and the China Scholarship Council (CSC). The earlier version of this paper was presented at the 16th International Schumpeter Society Conference held in Montreal between July 6th and 8th, 2016.

## Appendix. Supplementary data

Supplementary data associated with this article includes 6 tables and would be available from [doi:10.17632/pf8t6s62zf.5](doi:10.17632/pf8t6s62zf.5).

Table S1: Citation fractions in 176 scientific categories for 4 decadal years arranged within the 16 super-disciplines;

Table S2: 176 scientific category vectors for each of the 44 technological domains;

Table S3: Top 10 most cited scientific categories for each of the 44 domains;

Table S4: Correlation matrix of the 44 domains based on the 176 scientific category vectors;

Table S5: 16 super-discipline vectors for 44 technological domains and the correlation matrix among the vectors for the 44 domains;

Table S6: Correlation matrix among the 44 domains based upon inter-domain citations of patents by patents.